\documentclass[aps,prl,twocolumn,showpacs,byrevtex]{revtex4}
\let\oldhat\hat
\renewcommand{\hat}[1]{\oldhat{\mathbf{#1}}}
\usepackage{times,mathptm}
\usepackage[dvips]{graphicx,color}
\usepackage{dcolumn}
\usepackage{makecell}

\begin{document}
\title{Density-functional calculations of multivalency-driven formation of Te-based monolayer materials with superior electronic and optical properties }
\author{Zhili Zhu$^{1}$, Xiaolin Cai$^{1}$, Chunyao Niu$^1$, Seho Yi$^2$, Zhengxiao Guo$^{3,1}$, Feng Liu$^4$, Jun-Hyung Cho$^{5,2,1}$, Yu Jia$^{1*}$,  and Zhenyu Zhang$^{5*}$   }
\affiliation{$^1$ International Laboratory for Quantum Functional Materials of Henan, and School of Physics and Engineering, Zhengzhou University, Zhengzhou 450001, China \\
$^2$ Department of Physics, Hanyang University, 17 Haengdang-Dong, Seongdong-Ku, Seoul 133-791, Korea \\
$^3$ Department of Chemistry, University College London, London WC1E 6BT, United Kingdom \\
$^4$ Department of Materials Science and Engineering, University of Utah, Salt Lake City, Utah 84112, USA \\
$^5$ ICQD, Hefei National Laboratory for Physical Sciences at the Microscale, and Synergetic Innovation Center of Quantum Information and Quantum Physics, University of Science and Technology of China, Hefei, Anhui 230026, China 
}
\date{\today}

\begin{abstract}
Contemporary science is witnessing a rapid expansion of the two-dimensional (2D) materials family, each member possessing intriguing emergent properties of fundamental and practical importance. Using the particle-swarm optimization method in combination with first-principles density functional theory calculations, here we predict a new category of 2D monolayers named tellurene, composed of the metalloid element Te, with stable 1T-MoS$_2$-like ( ${\alpha}$-Te), and metastable tetragonal (${\beta}$-Te) and 2H-MoS$_2$-like (${\gamma}$-Te)  structures. The underlying formation mechanism of such tri-layer arrangements is uniquely rooted in the multivalent nature of Te, with the central-layer Te behaving more metal-like (e.g., Mo), and the two outer layers more semiconductor-like (e.g., S). In particular, the ${\alpha}$-Te phase can be spontaneously obtained from the magic thicknesses truncated along the [001] direction of the trigonal structure of bulk Te. Furthermore, both the ${\alpha}$- and ${\beta}$-Te phases possess electron and hole mobilities much higher than MoS$_2$, as well as salient optical absorption properties. These findings effectively extend the realm of 2D materials to group-VI monolayers, and provide a new and generic formation mechanism for designing 2D materials.
\end{abstract}
\pacs{73.20.At, 61.46.-w, 73.22.-f, 73.61.Cw}

\maketitle

The two-dimensional (2D) materials have been intensively investigated in recent years for their intriguingly emergent properties that can be exploited for electronic, photonic, spintronic, and catalytic device applications~\cite{1,2,3,4,5,6,7,8,9,10}. Various 2D monolayers have been synthetized beyond the first member system of graphene~\cite{1,2,3}, including the group-IV monolayers of silicene~\cite{4} and stanene~\cite{8}, the group-V monolayer of phosphorene~\cite{5}, and the group-III monolayer of borophene~\cite{6,7}. Besides these group-III, -IV, and -V elemental monolayers, transition metal dichalcogenides (TMDCs) have also been attracted much attention because of their relatively wider, tunable, and direct band gaps and inherently stronger spin-orbit coupling~\cite{9,10}. Yet to date, somewhat surprisingly, no prediction or fabrication of group-VI elemental monolayers has been made, whose potential existence would not only further enrich our understanding of the realm of the 2D materials world, but could also offer new application potentials stemming from their uniquely physical and chemical properties.  
 
In this Letter, we add an attractive new category to the ever increasing 2D materials family by predicting the existence and fabrication of group-VI elemental monolayers centered on the metalloid element Te. Our theoretical calculations reveal that 2D monolayers of Te, named tellurene, can exist in the stable 1T-MoS$_2$-like ( ${\alpha}$-Te) structure, and metastable tetragonal (${\beta}$-Te) and 2H-MoS$_2$-like (${\gamma}$-Te) structures. These tri-layer arrangements are driven by  the unique multivalency nature of Te, with the central-layer Te behaving more metal-like, and the two outer layers more semiconductor-like. In particular, the monolayer and multilayers of ${\alpha}$-Te can be readily obtained via a thickness-dependent structural phase transition from the trigonal bulk Te, with van der Waals-type coupling between neighboring tri-layers. Furthermore, both the ${\alpha}$- and ${\beta}$-Te phases possess not only higher carrier mobilities ranging from hundreds to thousands of cm$^2$V$^{-1}$s$^{-1}$ compared to MoS$_2$, but also  significantly enhanced  optical absorption properties due to a nearly direct or direct band gap. These findings effectively extend the realm of 2D materials to group-VI monolayers, and provide a new and generic formation mechanism for designing 2D materials.

We perform the particle-swarm optimization (PSO) searches~\cite{11} in combination with the DFT calculations using the Vienna ab initio simulation package (VASP) within the projector augmented wave method~\cite{12,13}. For the exchange-correlation energy, we employ the PBE functional~\cite{14} with the van der Waals (vdW) correction proposed by Grimme~\cite{21} and the screened hybrid functional, HSE06, which can typically describe the band gaps better~\cite{15,16}. Unless otherwise specified, the Te monolayers are modeled by a periodic 1${\times}$1${\times}$1 slab geometry with a vacuum thickness of ∼20 {\AA}. The kinetic-energy cutoff for the plane wave basis set is chosen to be 500 eV, and the k-space integration is done using the Monkhorst-Pack scheme with the 21${\times}$21${\times}$1 meshes in the Brillouin zones. All the atoms are allowed to relax along the calculated forces of less than 0.01 eV/{\AA}. The phonon calculation is performed using larger supercells, as implemented in the Phonopy code~\cite{17}.

\begin{figure}[ht]
\includegraphics[width=8cm]{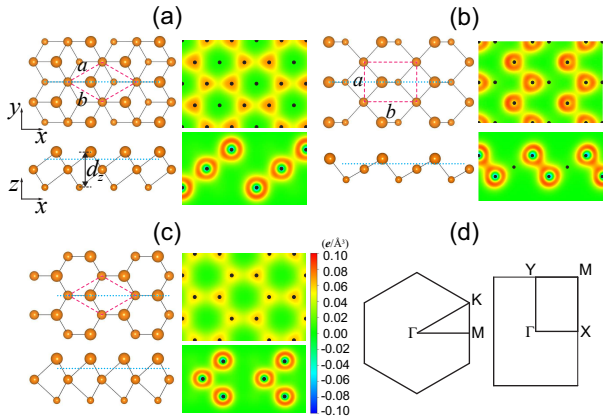}
\caption{(Color online) Top and side views of the optimized structures of tellurene in different phases: (a) ${\alpha}$-Te, (b) ${\beta}$-Te, and (c) ${\gamma}$-Te. The Brillouin zones for ${\alpha}$ (or ${\gamma}$) and ${\beta}$ phases are drawn in (d). The dashed line indicates the unit cell of each structure. For distinction, the large, medium, and small circles represent Te atoms located in the upper, central, and lower layers, respectively. The total charge density of each structure is plotted at the horizontal and vertical cross sections indicated by the blue dotted lines.}
\label{figure:1}
\end{figure}

Figure 1 presents the optimized structures of Te monolayers or tellurene. We identify three different phases denoted by ${\alpha}$-, ${\beta}$-, and ${\gamma}$-Te, as shown in Fig. 1(a), 1(b), and 1(c), respectively. The structural parameters and cohesive energy of each optimized structure are listed in Table I. It is seen that ${\alpha}$-Te has the 1T-MoS$_2$-like structure containing three Te atoms per unit cell. Here, when compared with 1T-MoS$_2$ monolayer, there are the two distinct types of Te atoms with different coordination numbers (\emph{$n_c$}): a central Te atom located at the Mo site has $n_c$ = 6, while a Te atom in the upper or lower layer at the S sites has $n_c$ = 3. Meanwhile, ${\beta}$-Te is composed of the planar four-membered and chair-like six-membered rings arranged alternately with the lattice constants $a$ = 4.17 and $b$ = 5.49 {\AA} [Table I]; in this structure, a central Te atom has $n_{c}$ = 4, while an upper or lower Te atom has $n_c$ = 3. ${\gamma}$-Te has the 2H-MoS$_2$-like structure, with smaller lattice constants $a$ = $b$ = 3.92 {\AA} than those ($a$ = $b$ = 4.15 {\AA}) of ${\alpha}$-Te. Correspondingly, the bond length ($d$ = 3.08 {\AA}) and interval distance ($d_z$ = 4.16 {\AA}) between the upper and lower Te atoms in ${\gamma}$-Te are larger than those ($d$ = 3.02 {\AA} and $d_z$ = 3.67 {\AA}) in ${\alpha}$-Te. To examine the relative stability of different tellurene allotropes, the cohesive energy ($E_c$) per atom with respect to the energy of an isolated Te atom is calculated. According to the results in Table1, ${\alpha}$-Te is energetically the most stable phase, while ${\beta}$-, and ${\gamma}$-Te are the meta-stable phases.

To examine the structural stability of tellurene, we perform the phonon calculations, which can identify the potential presence of soft phonon modes that may lead to structure instability. The calculated phonon spectra of tellurene are displayed in Fig. S1 of the Supplemental Material. We first confirm that all he phases are thermodynamically stable without imaginary-frequency phonon modes. The dynamic stability is further investigated using \emph{ab initio} molecular dynamics simulations. We find that the equilibrium structures of ${\alpha}$-Te and ${\beta}$-Te hardly change at room temperature, while ${\gamma}$-Te becomes unstable at temperatures above $\sim$200K. In the movies of the Supplemental Material, we illustrate the dynamic stability of each phase at 300 K up to a time period of 3 \emph{ps} with 1 \emph{fs} time step.

\begin{table}[ht]
\caption{Structural parameters of tellurene, together with the cohesive energy $E_c$ and the charge transfer ${\Delta}$Q from the central atom to the ourter atoms: $a$ and $b$ are the lattice constants, $d$ is the bond length, and $d_z$ is the interval distance between the upper and lower Te layers (see Fig. 1). For comparison, the structural and energetic properties of bulk Te are also listed. }
\begin{ruledtabular}
{\renewcommand{\arraystretch}{2.0}
\begin{tabular}{cccccc}
   & $a$, $b$ ({\AA}) & $d$ ({\AA}) & $d_z$ ({\AA}) & $E_c$ (eV/atom)& ${\Delta}$Q ($e$) \\\hline
 ${\alpha}$-Te & $a$ = $b$ = 4.15 & 3.02 & 3.67 & 2.62 & 0.41 \\
 ${\beta}$-Te  & \makecell{$a$ = 4.17 \\ $b$ = 5.49}    & \makecell{3.02 \\ 2.75\footnotemark[1]}& 2.16 & 2.56 & 0.11 \\
 ${\gamma}$-Te & $a$ = $b$ = 3.92 & 3.08 & 4.16 & 2.46 & 0.29 \\ 
{Te-I} &\makecell{$a$ = $b$ = 4.33 \\ $c$ = 6.05}& 2.90 & -- & 2.79 & -- \\
\end{tabular}
}
\end{ruledtabular}
\footnotetext[1]{bond length between two Te atoms with $n_c$ = 3.}
\label{table:1}
\end{table}

In Figs. 1(a)-1(c), the total charge densities of ${\alpha}$-, ${\beta}$-, and ${\gamma}$-Te reveal their bonding characteristics, respectively. For ${\alpha}$- and ${\gamma}$-Te, there exists a metal-ligand-like bonding between the central atom and the outer atoms. On the other hand, for ${\beta}$-Te, the outer atoms with are bonded to each other with the ${\sigma}$ bond, while the central atoms interact with the outer atoms in the form of a metal-ligand-like bonding. Consequently, the former bond length (2.75 {\AA}) is much shorter than the latter one (3.02 {\AA}). According to the Bader charge analysis, the charge transfer from the central to the outer atoms amounts to 0.41, 0.11, and 0.29$e$ in ${\alpha}$-, ${\beta}$-, and ${\gamma}$-Te, respectively (see Table I), and therefore the central Te atoms behave more metal-like with a larger $n_c$ while the outer Te atoms more semiconducting with a smaller $n_c$. The structural features of tellurene can be further associated with the bonding characters of group-VI elements, where the nonmetallic character is weakened in the order of O $>$ S $>$ Se $>$ Te, leading to a complete metallic character of Po. In particular, Te has the dual characteristics of both metal and nonmetal. It is thus feasible that the two dimensional monolayers of Te can adopt the tri-layer atomic structures, , e.g. MoS$_2$-like structure. With the dimensionality reduction, the multivalency-dominated 2D structures with heterogeneous coordination numbers become lower in energy. Collectively, these findings amply reflect the distinct multivalent nature of Te and its vital role in the formation of tellurene. 

\begin{figure}[ht]
\includegraphics[width=8.5cm]{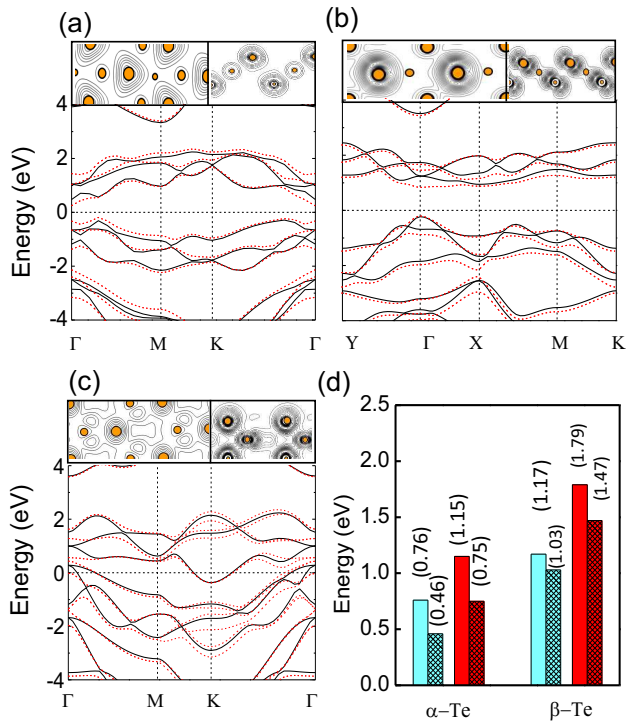}
\caption{(Color online) Band structures of (a) ${\alpha}$-Te, (b) ${\beta}$-Te, and (c) ${\gamma}$-Te, obtained using the PBE scheme without  (solid) and with (dashed) inclusion of the SOC. The contour plots of the electron densities for the valence states within 0.5 eV below the valence band maximum or Fermi level $E_F$ are drawn on the same horizontal and vertical cross sections as marked in Fig. 1. The interval of the charge contours is 1${\times}$10$^{-3}$ electrons/{\AA}$^3$. (d) Band gaps obtained using the PBE (PBE+SOC) and HSE (HSE+SOC), as represented by the cyan (cyan meshed) and red (red meshed) bars, respectively.}
\label{figure:2}
\end{figure}

Figures 2(a)-2(c) show the band structures of ${\alpha}$-, ${\beta}$-, and ${\gamma}$-Te, respectively, obtained using the PBE calculation. We find that ${\alpha}$- and ${\beta}$-Te are semiconductors with indirect band gaps of $E_g$ = 0.76 and 1.17 eV, respectively, while ${\gamma}$-Te is a metal. It is well-known that the semi-local PBE scheme underestimates the band gap. In order to remedy such a deficiency in PBE, we perform the hybrid DFT calculation with the HSE06 functional, which is known to provide better predictions of the band gaps. As shown in Fig. 2(d), the HSE calculations for ${\alpha}$- and ${\beta}$-Te give increased $E_g$ = 1.15 and 1.79 eV, respectively. Given the heavy metal nature of Te, we also examine the effects of SOC on the band structure. The results obtained using the PBE+SOC calculation are plotted with the dashed lines in Figs. 2(a)-2(c). We find that the inclusion of SOC in ${\alpha}$- and ${\beta}$-Te induces a transformation from an indirect to a nearly direct and a direct band gap at the ${\Gamma}$ point, respectively. This indirect-to-direct band-gap change in ${\alpha}$- and ${\beta}$-Te may significantly enhance their optical absorbance. Indeed, as seen in Fig. 3,  both ${\alpha}$- and ${\beta}$-Te exhibit superb optical absorptions, which can be exploited for optoelectronics and photon detection. ${\beta}$-Te also exhibits optical anisotropies, with stronger absorbance along the zigzag chain direction, which can be exploited for developing polarized optical sensors. Furthermore, it is noted that the PBE+SOC (HSE+SOC) band gaps of ${\alpha}$- and ${\beta}$-Te are reduced by 0.30 (0.40) and 0.26 (0.32) eV, compared to the PBE (HSE) ones: see Fig. 2(d). Therefore, the HSE+SOC band gap of ${\alpha}$-Te becomes 0.75 eV, which is located between the band gaps (${\sim}$0.7 and ${\sim}$1.1 eV) of bulk Ge and Si~\cite{18}, and that of ${\beta}$-Te is 1.47 eV, which is close to that of GaAs. These physically realistic values of the band gaps of the stable and meta-stable tellurene phases may offer desirable (e.g., ohmic) contacts when such materials are integrated for device applications.

\begin{figure}[ht]
\includegraphics[width=8.5cm]{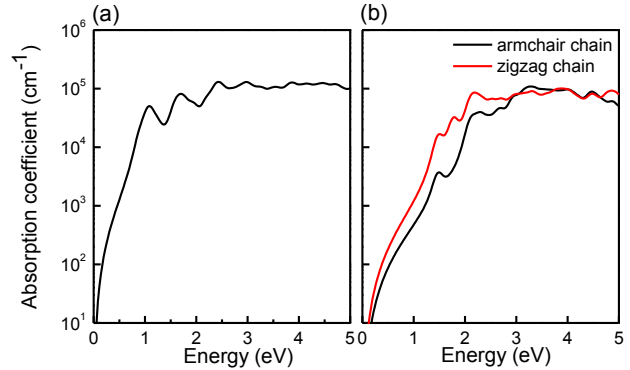}
\caption{(Color online) Calculated optical absorption coefficients for (a) ${\alpha}$-Te, (b)${\beta}$-Te. In (b), the polarization is along the zigzag or armchair chain direction.}
\label{figure:3}
\end{figure}

For potential technological applications in electronic devices, the newly discovered 2D materials should have sufficiently high carrier mobilities. To estimate the carrier mobility of the tellurene monolayers, we calculate their effective masses, which are relatively smaller than those ($m^*_e$= 0.47, and $m^*_h$= 0.58 $m_e$) of monolayer 2H-MoS$_2$ (see Table II). These results suggest that tellurene may possess high electron and hole mobilities. Using the acoustic phonon limited method~\cite{19}, the room-temperature carrier mobilities of ${\alpha}$- and ${\beta}$-Te are found to range from hundreds to thousands of cm$^2$V$^{-1}$s$^{-1}$, much higher than those of monolayer 2H-MoS$_2$(see Table II and the Supplemental Material). Here, ${\mu}_e$ and ${\mu}_h$ show large differences in magnitude, indicating asymmetric mobilities of electrons and holes due to their different effective masses. In addition, ${\beta}$-Te has anisotropic characters of electron and hole mobilities along the $y$ direction.

\begin{table}[h]
\caption{Effective masses $m^*$ and carrier mobilities ${\mu}$ of ${\alpha}$-Te, ${\beta}$-Te, and 2H-MoS$_2$, obtained using the PBE+SOC calculation. For the tetragonal structure of ${\beta}$-Te, the components along the armchair chain ($x$) and zigzag chain ($y$) directions are separately given.}
\begin{ruledtabular}
{\renewcommand{\arraystretch}{2.0}
\begin{tabular}{ccccc}
 &\multicolumn{2}{c}{$m^*$ ($m_e$)} &\multicolumn{2}{c}{${\mu}$ ($10^3$ cm$^2$V$^{-1}$s$^{-1}$)}\\ \hline
             & electron & hole & electron & hole \\
${\alpha}$-Te& 0.11& 0.17& 2.09& 1.76 \\
${\beta}$-Te & \makecell{0.83 ($x$) \\ 0.19 ($y$)} & \makecell{0.39 ($x$) \\ 0.11 ($y$)} & \makecell{0.05 ($x$) \\ 0.10 ($y$)} & \makecell{1.98 ($x$) \\ 0.45 ($y$)}\\
2H-MoS$_2$  & 0.47& 0.58 & 0.08& 0.29 \\

\end{tabular}
}
\end{ruledtabular}
\label{table:3}
\end{table}

Now, we turn to discuss possible fabrication route for tellurene and its multilayers. To date, the existing 2D materials can be divided into two categories: one that can be mechanically exfoliated from its layered bulk counterpart, such as graphene and MoS$_2$; the other, acking a layered bulk counterpart, has to be grown epitaxially on a  proper  substrate, such as silicene and stanene. Meanwhile, the Te-I bulk has the form of helical chains along the c axis, and the Te films most easily grow in the [001] direction~\cite{20}, totally different from the structure of tellurene. Surprisingly, the monolayer or multilayers of tellurene can be generated via the new formation mechanism characterized by a thickness-dependent structural phase transition in the ultrathin film regime, as discussed below.

We reach the above important finding through a systematic study of the Te film stability with increasing film thickness, determined by the formation energy ($E_f$) as a function of the number of atomic layers, \emph{N}. Here the initial configurations of the slabs are taken by truncating the trigonal structure of bulk Te (hereafter termed Te-I) along the (001) direction. The formation energy is given by the cohesive energy difference $E_f = (E_{slab}(\emph{N})-\emph{N}E_{bulk})/\emph{N}$, where $E_{slab}(\emph{N})$ and $E_{bulk}$ are the total energies of the slab and a single layer in bulk Te, respectively. For these multilayerd systems, we have included the vdW interactions using the DFT-D2 method~\cite{21}. 

Figure 4(a) shows the formation energy variations of the fully relaxed Te slabs with increasing \emph{N}, exhibiting a distinct oscillatory behavior. There are five highly preferred (or magic) thicknesses of \emph{N} = 3, 6, 9, 12, 15 when the thickness of the Te-I slabs increases from \emph{N} = 1 to 20. Strikingly, we find that the Te slabs automatically transform into multilayered structures of α-Te at the magic layer thicknesses, while the Te slabs will keep the chain-like structures of bulk Te away from these magic thicknesses. The insert in Fig. 4(a) highlights the stability of the different slabs by the second difference, while Fig. 4(c) and (d) highlight the different structural preferences of Te slabs with \emph{N} = 8 and 9, respectively. We further obtain that the interlayer coupling strength between two neighboring tellurene monolayers (or, equivalently, two Te trilayers) of Te is 26 meV/{\AA}$^2$, which is on the same order as that of  MoS$_2$ (21 meV/{\AA}$^2$ )~\cite{22}, suggesting that a single tellurene layer can be readily exfoliated once it is formed.

\begin{figure}[ht]
\includegraphics[width=8cm]{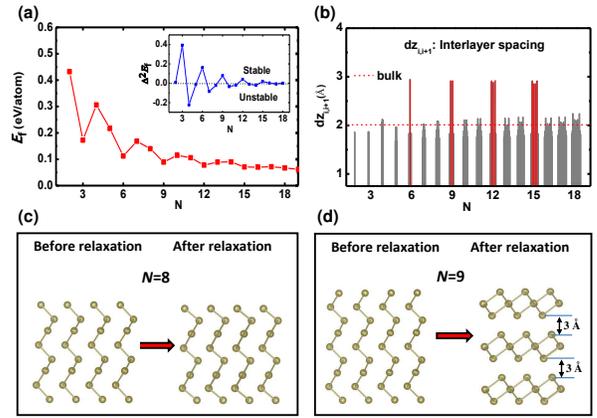}
\caption{(Color online) Formation energies, stabilities, and interlayer spacings of Te slabs at different thicknesses. (a) Formation energies of the fully relaxed Te slabs as a function of thickness. The insert represents the second-order difference of $E_f$, with positive values indicating stable systems. (b) Distribution of the layer-resolved interlayer spacing of relaxed Te slabs as a function of thickness. The dashed line denotes the atomic layer spacing of bulk Te. (c) and (d) are the geometric structures of Te slabs at \emph{N} = 8 and 9 before and after structural optimization, respectively. } 
\label{figure:4}
\end{figure}

At present, there is no a priori knowledge about which vdW scheme is more accurate for a given system. As cross checks, we have also examined the Te film stability as a function of the film thickness with the vdW interactions treated within the widely adopted first-principles-based schemes of Tkatchenko and Scheffler~\cite{23} (DFT-TS) and vdW-DF2~\cite{24}, respectively. For either scheme, the results qualitatively also support the existence of a structural phase transition from the bulk-truncated Te structure to multilayered tellurene at the identical film thicknesses, but the number of such magic thicknesses is varied depending on the specific version of the vdW scheme. For DFT-TS (vdW-DF2), the layered (or close shelled) structures are found to be highly preferred at the thicknesses of \emph{N} = 3, 6, 9 (3, 6). Here, we note that the optimized lattice parameters of bulk Te obtained using the vdW-DF2 scheme show more severe deviation from the experimental values (see Table S1 of the Supplemental Material), while the DFT-D2 and DFT-TS schemes agree better with experiments. Together, these results convincingly indicate that at least a few monolayers of the 2D tellurene structure will be readily obtained in a typical yet thickness-controlled fabrication approach on a proper substrate favoring layered growth. 

In summary, our state-of-the-art global structural searching combined with first-principles calculations has resulted in the discovery of a new category of 2D materials composed of the group-VI element of Te. These new 2D materials called tellurene can be stabilized in the MoS$_2$-like (${\alpha}$-, ${\gamma}$-Te) or tetragonal (${\beta}$-Te) structures, and their underlying formation mechanism is inherently rooted in the multivalency nature of Te. The ${\alpha}$-Te and ${\beta}$-Te monolayers not only exhibit superb optical absorptions, but also possess much higher carrier mobility than MoS2. The ${\alpha}$-Te multilayers can be achieved spontaneously from the bulk truncated films via a novel thickness-dependent structural phase transition. The coupling between neighboring tellurene layers is of vdW type, allowing easy separation of a tellurene layer via mechanical exfoliation. The superior electronic and optical properties of tellurene are expected to find broad technological applications.

\vspace{0.4cm}

We thank Dr. Xiaoyu Han and Prof. Qiang Sun for helpful discussions. This work was partially supported by the NSFC (Nos. 11274280, 11504332, 11634011, 61434002), the National Basic Research Program of China (Nos. 2012CB921300 and 2014CB921103). Z.X.G. is supported by the UK EPSRC (No. EP/K021192/1). J.-H.C. is supported by the National Research Foundation of Korea (NRF) grant funded by the Korea Government (No. 2015M3D1A1070639). F.L. is supported by U.S. DOEBES (No. DE-FG02-04ER46148).

\vspace{0.4cm}

\noindent $^{*}$ Corresponding authors: jiayu@zzu.edu.cn, zhangzy@ustc.edu.cn.


\end{document}